\documentclass[twocolumn,preprintnumbers,amsmath,amssymb,superscriptaddress]{revtex4}

\usepackage{graphicx}
\usepackage{color}
\usepackage{dcolumn}
\usepackage{bm}

\begin{document}

\author{William N. Plick}
\affiliation{University of Dayton, Department of Physics, 300 College Park, Dayton, OH, 45469, United States}

%\affiliation{LTCI, CNRS, Departement Informatique et Reseaux, Telecom ParisTech,
%23 Avenue d'Italie, CS 51327, 75214 Paris CEDEX 13, France}

\author{Francesco Arzani}

\author{Nicolas Treps}
\affiliation{Laboratoire  Kastler  Brossel,  Sorbonne  Universit\'{e}s,  CNRS,  ENS-PSL  Research  University,
Coll\'{e}ge  de  France,  CNRS;  4  place  Jussieu,  F-75252  Paris,  France}

\author{Eleni Diamanti}
\affiliation{Laboratoire  d'Informatique  de  Paris  6,  CNRS, Sorbonne  Universit\'{e},  4  place  Jussieu,  75005  Paris, France}

\author{Damian Markham}
\affiliation{Laboratoire  d'Informatique  de  Paris  6,  CNRS, Sorbonne  Universit\'{e},  4  place  Jussieu,  75005  Paris, France}

\title{Violating Bell inequalities with entangled optical frequency combs and multi-pixel homodyne detection}
\date{\today}

\begin{abstract}
\noindent We have theoretically investigated the possibility of using any of several continuous-variable Bell-type inequalities \--- for which the dichotomic measurements are achieved with coarse-grained quadrature (homodyne) measurements \--- in a multi-party configuration where each participant is given a section, in the frequency domain, of the output of an optical parametric oscillator which has been synchronously-pumped with a frequency comb. Such light sources are undergoing intense study due to their novel properties, including the potential for production of light entangled in many hundreds of physical modes \--- a critical component for many proposals in optical or hybrid-optical quantum computation proposals. The situation we study notably uses only highly-efficient optical homodyne detection, meaning that in such systems the fair-sampling loophole would be relatively easy to avoid.            
\end{abstract}

\maketitle

\section{Introduction}

Bell's inequalities allow us to unambiguously (under certain reasonable assumptions) divide phenomena into two types: those that are describable with local-realistic probability theories, and those that must be described with the wider probability space of quantum mechanics \cite{BellNL}. This is important not only because of its deep philosophical implications, but also because systems that behave quantum-mechanically may be utilized for technologies in-principle more powerful than their classical counterparts, such as metrology \cite{met,met2}, cryptography \cite{crypt}, and computation \cite{qc}.    

Of the several approaches to quantum computing, one that has received a significant amount of attention is the so-called ``measurement-based'' or ``cluster-state'' approach \cite{MBQC}. 

Measurement-based quantum computing (MBQC) is an alternative view to the standard ``circuit model'' of quantum computing \--- where the computation is envisioned as a set of quantum-logical gates enacted in sequence, some of which generate entanglement. In MBQC a very large quantum state is created initially, including all the entanglement that will be needed. A sequence of measurements is then performed where the results of previous measurements determine which measurements are performed next, the state at the end then encodes the answer to the desired computation. The two approaches are formally equivalent, but differ significantly in their implementation.     

A further modification of MBQC is found when we allow the modes that comprise the initial, highly-entangled, state to take continuous-variable values \cite{CVMBQC}, such as the $p$ and $q$ quadrature modes of an electromagnetic field. We will consider such fields in the optical domain in this paper. Continuous-variable MBQC (CV-MBQC) relies on the ability to generate very large, highly-entangled states between thousands or millions of physical modes. 

This is a difficult task but a promising approach is to use a device called a ``synchronously-pumped optical parametric oscillator'' (SPOPO) to generate the required states. Such a device takes the output of a frequency comb and uses it as the pump for an optical parametric oscillator. This process, in principle, entangles every frequency ``tooth'' of the comb with every other \--- to varying degrees. See, for example, Refs.\cite{Fabre1,pfister1,Fabre2,furusawa,pfister2,huntington}.

This device can also be modified such that one or more photons may be subtracted off coherently from any combination of frequency modes via a quantum pulse gate (QPG) \cite{qpg1,qpg2,qpg3}. In those frequency modes from which a photon has been subtracted the Wigner-function representation gains negative regions \--- an indicator of non-classicality and the potential to use these modes as resources in quantum-enabled tasks in information technology such as quantum computing and key distribution.

Furthermore, a portion of the initial frequency comb (pump) may be partitioned off for later use as a local oscillator. This local oscillator can be shaped arbitrarily in the frequency domain, permitting the homodyne detection of, in principle, any mode which has a pure-state representation in the frequency basis.

Homodyne detection has the advantage that it is highly efficient and thus if the goal is to violate a Bell-type inequality over a lossy or \--- more poignantly \--- untrusted channel, such detection is advantageous for avoiding or closing the fair-sampling (efficiency) loophole.

In the system we consider here the final detection (and homodyning) is done across a detector with many individual pixels after another diffraction grating. Therefore the different sections of the frequency domain are coarse grained into the desired number of parties, each of which is given a section of the full frequency range.        

In principle each party, as defined above, possesses at the moment before detection, a state which is entangled with all other parties in a complicated manner. The goal of the investigation reported here is to analyze a generalized version of this system and determine the prospects for violating a continuous-variable Bell inequality, the strongest possible proof of non-classicality, for some number of parties.

We find that \--- for the two party case \--- the photon-subtracted SPOPO with homodyning is capable of a higher violation than has been shown previously in the analysis of simpler systems. Though this is promising it is still likely below what could be distinguished experimentally (high efficiency is no guarantee of a high signal-to-noise). We also analyze the four-party case. What we find is a massive parameter space which is, at least, very computationally intensive to search. We consider the simplest and most direct cases and find that no violation is possible. We stress that this is merely in the sections of that space that are easily searchable, and that violations are likely in other regions, though determining which regions these are is highly non-trivial. 

We plan to make all models and programs publicly available in the hopes that this stimulates a more through investigation.

The remainder of this manuscript is organized as follows: Section II  reviews synchronously pumped optical parametric oscillators in general. In Section III we synthesize the setup in total, creating a mathematical model. Section IV reviews the kind of local-realistic inequality we use, and how discrete observables can be constructed from the continuous measurements available in homodyne detection. In Section V we examine the two-party case, and in VI the four-party. Section VII is left for discussions, conclusions, and a review of the results.  

\section{The Synchronously Pumped Optical Parametric Oscillator (SPOPO)}

The ability to scale up to architectures of useful size is a key requirement for quantum information applications. Experimentally, systems with many optical modes can be realized with optical frequency combs. These are in turn provided by mode-locked lasers, whose output is a train of identical light pulses in the temporal domain. The spectrum of such a field, obtained by Fourier-transform, consists of a series of on the order of millions of equally-spaced, single-frequency modes, namely, the frequency comb. Each ``tooth'' of the comb can be used as an independent quantum mode, allowing for the creation of a state on the order of a million modes. 

However in order to couple these modes and \--- hopefully \--- create entanglement between them for use as a resource in quantum technologies they must be made to interact. 

There are many variations on this approach, but the basic principle is to inject the output of the comb into an optical parametric oscillator (OPO \--- an optical cavity containing a non-linearity) which has the same mode structure as the comb. For full details on the various approaches see \cite{patera,ferrini,Fabre1,Fabre2,furusawa,pfister2,trepsnat,witness,opt}.

Mathematically, these approaches can be modeled as an interaction Hamiltonian working on all the various modes, commonly called the joint spectral distribution (or joint spectral amplitude) and written as             

\begin{eqnarray}
\hat{H}_{I}=\sum_{mq}\mathcal{L}_{mq}\hat{a}_{m}\hat{a}_{q}+\mathrm{h.c.},\label{ham}
\end{eqnarray}

\noindent where the $\hat{a}$'s are the annihilation operators working on (corresponding to) the modes represented by the index. The physical nature of the modes is left undefined for now. The matrix $\mathcal{L}$ is the coupling between these modes. Determining this matrix for our specific implementation is discussed in Section V. For the setup we model in the following sections of this paper there are 3 million individual modes. 

The relative squeezing induced between modes can be different for each combination of modes, leading to a relatively sticky mathematical situation. Things are simplified greatly by considering the ``supermode basis''. The supermode basis is constructed of the modes which diagonalize the coupling matrix $\mathcal{L}_{mq}$. They are found by determining the eigensystem of $\mathcal{L}_{mq}$ (NB: This is only true when the coupling matrix is real. In the general case the diagonalization is done by congruence \--- Autonne-Takagi factorization \--- a special case of singular value decomposition. If $\mathcal{L}$ is real, these coincide with eigenvalues, but in the general case a different procedure is needed, see \cite{AT}). Each eigenvector represents a conversion between the new supermode basis and the basis of the optical frequency (the comb's teeth). The eigenvalue associated with each eigenvector is the single-mode squeezing of that supermode. Since the matrix is diagonal there is no multi-mode squeezing in this representation. The supermodes are themselves entangled modes.    

\section{Setup Under Consideration}

\begin{figure}
\centering
\includegraphics[width=1\textwidth]{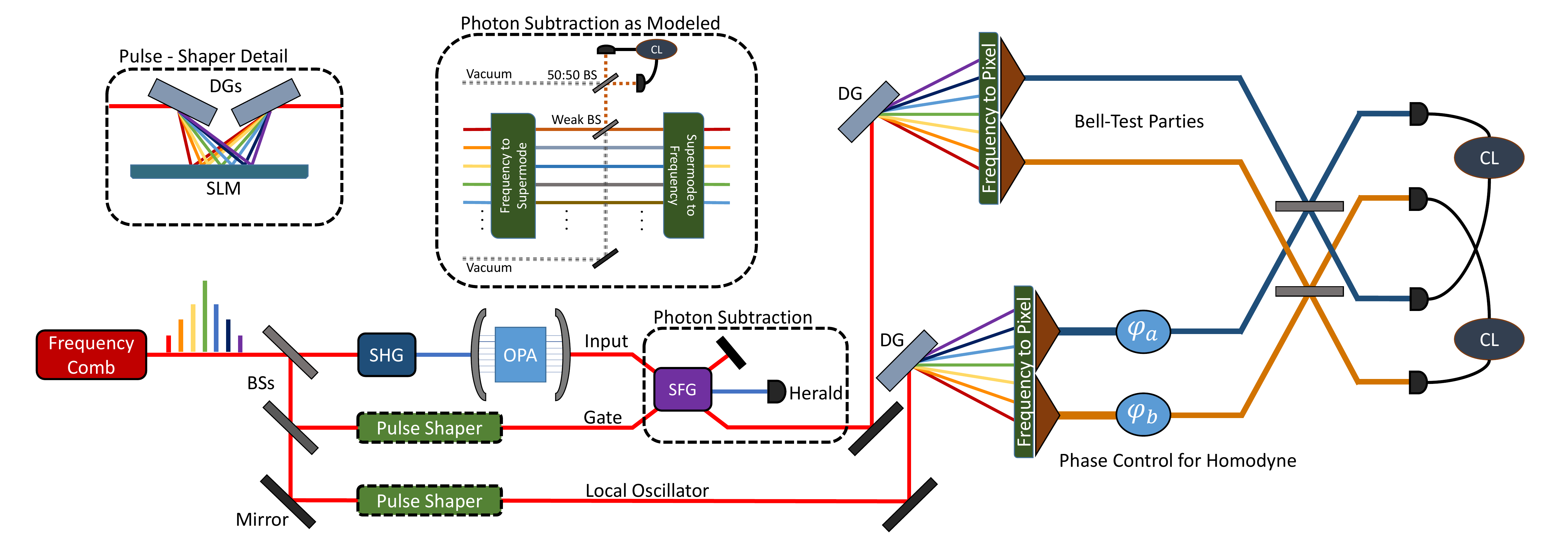}
\caption{A frequency comb is used as the pump in an optical parametric oscillator. Before the OPO a portion of the pump is shunted off with a beam-splitter to be pulse-shaped by two independent pulse shapers. Each pulse shaper consists of a diffraction grating, a spatial light modulator, and a second diffraction grating. One of these pulse-shaped beams is then used as a quantum pulse gate in a photon subtraction stage. We, however, mathemtically model this as weak beam splitter acting on the first super mode. If this beam splitter reflects two photons, and these are detected by a combination of a 50-50 beam splitter, on-off detectors and a coincidence logic (CL), then a two-photon subtracted state is heralded. The second pulse-shaped beam is prepared as a homodyning reference. The photon-subtracted SPOPO mode is then shone on a diffraction grating which fans the light out onto a detector with a resolution of between two and sixteen pixels. Each pixel mode of the SPOPO output is paired with an equivalent segment of the homodyne pulse. These modes are then coarse-grained in the same manner as the SPOPO mode and then each of these resultant modes is subjected to a phase shift as dictated by the desired local-realistic inequality and detection protocol. In this figure we specifically show the two-party case. Generalization to higher numbers of parties is straightforward. \label{setup}}
\end{figure}

The experimental setup we consider for theoretical study consists of a combination of many optical devices. The whole contraption is depicted in Figure \ref{setup}. A frequency comb is used as the pump in an optical parametric oscillator. Before the OPO a portion of the pump is shunted off with a beam-splitter to be pulse-shaped by two independent pulse shapers. Each pulse shaper consists of a diffraction grating (which maps the frequency domain to the spatial domain), a spatial light modulator (which then applies a pre-programmed phase and amplitude modulation to each segment of the frequency domain, thus shaping the pulse as a whole in an arbitrary and controllable manner), and a second diffraction grating which erases the position information and reconstructs the pulse.

One of these pulse-shaped beams is then used as a ``gate'' in a photon subtraction stage. The output of the SPOPO is fed into one of the input modes of a sum-frequency-generation crystal, while the gate beam is fed

\newpage

\onecolumngrid

\twocolumngrid

\noindent into the other. The detection of an up-converted photon in the heralding mode after the crystal then indicates that a photon was subtracted from both modes. Significantly, it can be guaranteed that the photon was subtracted \emph{coherently} from the combination of frequency modes present in the gate beam \--- therefore a photon may be subtracted from the SPOPO pulse in any mode including those of the supermode basis. For more information see Refs.\cite{qpg1,qpg2,qpg3}.

Since, when viewed in the supermode basis, the first supermode exhibits the largest squeezing (this is discussed in Section V and shown in Fig. 4), we subtract photons from that mode exclusively. This is done because photon subtraction creates more non-classicality (i.e. more negativity in the Wigner function) when subtracted from a more tightly-squeezed state).

A full discussion of the photon-subtraction procedure is beyond the scope of this paper. In fact we will model the subtraction as a very weak beam splitter and a photo-detection \--- a fairly standard method. In our case the ``beam splitter'' is modeled as working only on the first supermode. We also employ a second 50-50 beam splitter and a coincidence detection in order to model the subtraction of two photons. A two-photon subtraction is chosen since the parity of the number of modes must match the parity of the number of photons subtracted \cite{Acin}, and subtracting four (or more) photons is currently technologically unfeasible. 

The second pulse-shaped beam is then prepared as a homodying reference for continuous-variable quadrature measurements. With pulse shaping many combinations of frequency profiles are possible. We take a Gaussian distribution as this is what is used experimentally.

The photon-subtracted SPOPO mode is then shone on a diffraction grating which fans the light out onto a detector with a resolution of between two and sixteen pixels. The number of pixels represents the number of parties involved in the Bell test. Each party receives a segment of the frequency domain, the size of which is inversely proportionate to the number of parties. Those frequency modes in each segment are then all coarse-grained into a ``pixel mode'' which represents the signal each party corresponding to that pixel receives. It is likely that much coherence is lost at this stage. It should be noted that, of course, it would be impossible to close the locality loophole with the ``distant parties'' present on the same small device, but in principle these signals can be transmitted to distant locations as needed before measurement. In this paper we consider two possibilities: two parties, and four parties; though any other number is principle possible \--- the only limitations being the resolution of the detector and the quality of the diffraction grating.   

Each pixel mode of the SPOPO output is paired with an equivalent segment of the homodyne pulse. These modes are then coarse-grained in the same manner as the SPOPO mode and then each of these resultant modes is subjected to a phase shift as dictated by the desired local-realistic inequality and detection protocol. This is detailed in the following section.\\

\section{Continuous-Variable Bell-type inequalities with Photon-subtracted gaussian states and homodyne detection}  

We wish to utilize continuous (and Gaussian) homodyne measurements in the setup we examine for Bell-inequality violation due to their relative ease of application and high efficiency.  This is advantageous for closing (or avoiding) the detection (fair sampling) loophole and is of strong interest in quantum foundations and technologies. This was also first pointed out in Refs.\cite{raul,raulL}. However Bell inequalities are most commonly and directly constructed in terms of discrete variables. Therefore the first step of our analysis will be to discretize quadrature space so that it can be thought of as a dichotomic (two-valued) function. We follow the same procedure as in Refs.\cite{raul,raulL}.

\begin{figure}[h!]
\centering
\includegraphics[scale=0.7]{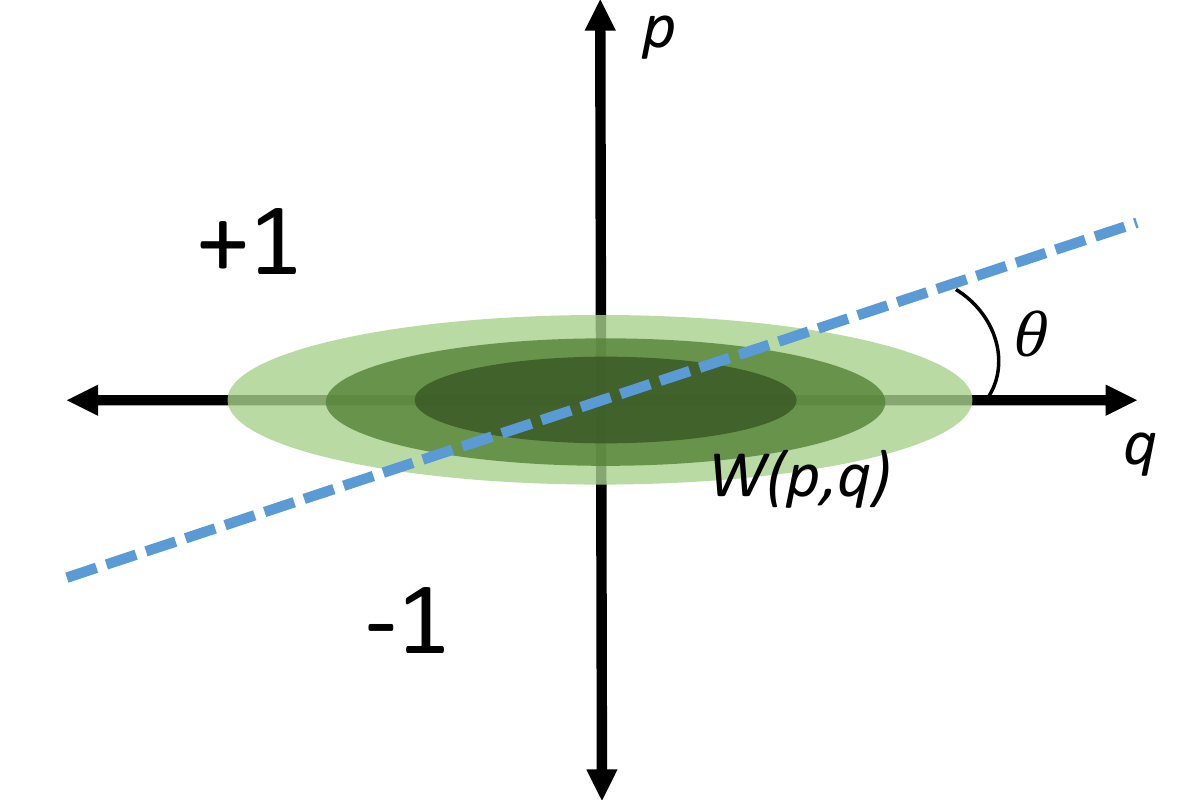}
\caption{A quadrature space with a Wigner function ($W$) living in it is split into two regions defined by the angle $\theta$. Half of the space is assigned the dichotomic value $+1$, and the other the value $-1$.  \label{dichot}}
\end{figure}

Given a particular choice of measurement angle ($\theta$) we draw a line through the origin of quadrature space such that the line is at angle $\theta$ to the horizontal, dividing the space into two regions. When a detection is made it will be to either side of this line. One side is assigned the value of $+1$ and the other a value of $-1$, and then the construction of the various Bell's inequalities proceeds normally. This is graphically depicted in Figure \ref{dichot}.

It is important to point out that the homodyne detection procedure is not inherently non-classical (i.e. it possesses a Gaussian Wigner function that is everywhere positive) so the quantum behavior is introduced at the stage of the photon subtractions. This is the opposite of most Bell tests where squeezed states (keeping in mind that spontaneous parametric down-conversion is just weak squeezing) are subjected to number resolving (or on-off) detection which is non-classical. Some non-classical element is needed in order to violate a Bell inequality otherwise the everywhere-positive Wigner distribution can be interpreted as precisely the hidden variable model such tests wish to disprove. 

We are further motivated to apply such the procedure outlined in Refs.\cite{raul,raulL} for the case of the photon-subtrated SPOPO due to some results found in Ref.\cite{Acin} that show that in the many-party generalization of such inequalities the amount of violation found scales with number of modes/parties present, as long as the number of modes is not of odd-parity. Since the SPOPO generates entanglement between very-many modes it is reasonable to assume that it would naturally lend itself to such a situation. This is the motivation for the work presented in this paper.    

For the two-party case, which we consider in the following section, the most obvious and direct inequality to apply is simply CHSH \cite{chsh}: 

\begin{eqnarray}
\mathcal{S}_{2}=\frac{1}{2}\left| E(a,b)+E(a',b)+E(a,b')-E(a',b')\right|\leq 1\label{chsh},
\end{eqnarray}     

\noindent where the $E$'s are the correlators and are defined in general as

\begin{eqnarray}
E(a,b)=\frac{R_{++}(a,b)+R_{--}(a,b)-R_{+-}(a,b)-R_{-+}(a,b)}{R_{++}(a,b)+R_{--}(a,b)+R_{+-}(a,b)+R_{-+}(a,b)}\nonumber
\end{eqnarray}  

where the $R$'s represent the rates of the events indicated by their subscripts as a function of the chosen measurement angles for Alice and Bob (i.e. $R_{+-}(a,b)$ is the rate of Alice measuring $+1$ when Bob measures $-1$ for settings $a$ and $b$ respectively). The primed and un-primed settings represent the two different choices each party uses.

In the next section we develop the mathematical formalism needed to check for violation of this inequality.\\

\section{Two-Party Case}

In this section we will study the simplest case available, which is when we limit ourselves to having the detector divided up into two regions (i.e. two pixel modes or parties). We will also take the simplest realistic parameters for other elements of the setup.

Note that all numerical values are taken \emph{directly} from the experiment described in the References: \cite{Fabre2,trepsnat,witness}. So our calculations can be said to closely model what is done experimentally.

We will draw a lot from Ref.\cite{patera} and from Ref.\cite{raul}. We follow the latter fairly closely. In those papers the authors consider a two-mode squeezed vacuum state, on each of which a photon subtraction is performed, followed by a homodyne detection on each mode. Essentially our work is an expansion of that formalism to a much more complicated physical scenario.

Much of the actual calculation is done in Mathematica and we have made a commented version of this available at Ref.\cite{math}. It should be noted that some of the calculations take a very long time on standard personal computing equipment so we also make available some pre-computed examples. All the aforementioned experimental values and parameters used in the calculation (or ranges of those values for parameters that may be varied) are given in that file. 

We start with the description of the relative coupling strengths of SPOPO modes as given in Ref.\cite{patera}:

\begin{eqnarray}
\mathcal{L}_{mq}=f_{mq}\alpha_{m+q},
\end{eqnarray}

\noindent where $f_{mq}$ is the phase-matching of the crystal for frequency modes $m$ and $q$, and is given by

\begin{eqnarray}
f_{mq}=\frac{\sin(\phi_{mq})}{\phi_{mq}}, \quad \phi_{mq}=\frac{1}{2}\left(k^{p}_{m+q}-k^{s}_{m}-k^{s}_{q}\right)l,
\end{eqnarray}

\noindent where $k^p$ and $k^s$ are the pump and signal wavenumbers, respectively; and $l$ is the length of the crystal. The factor $\alpha_{n}$ represents the amplitude and phase of the coherent pump at frequency $\omega_{n}$. 

The matrix $\mathcal{L}_{mq}$, the joint spectral distribution, representing the coupling between frequency modes $m$ and $q$, mediated by the crystal and pump. Thus the interaction Hamiltonian for the system is given as Eq.(\ref{ham}), which we repeat here: 

\begin{eqnarray}
\hat{H}_{I}=\sum_{mq}\mathcal{L}_{mq}\hat{a}_{m}\hat{a}_{q}+\mathrm{h.c.}
\end{eqnarray}   

\begin{figure}[h!]
\centering
\includegraphics[scale=0.37]{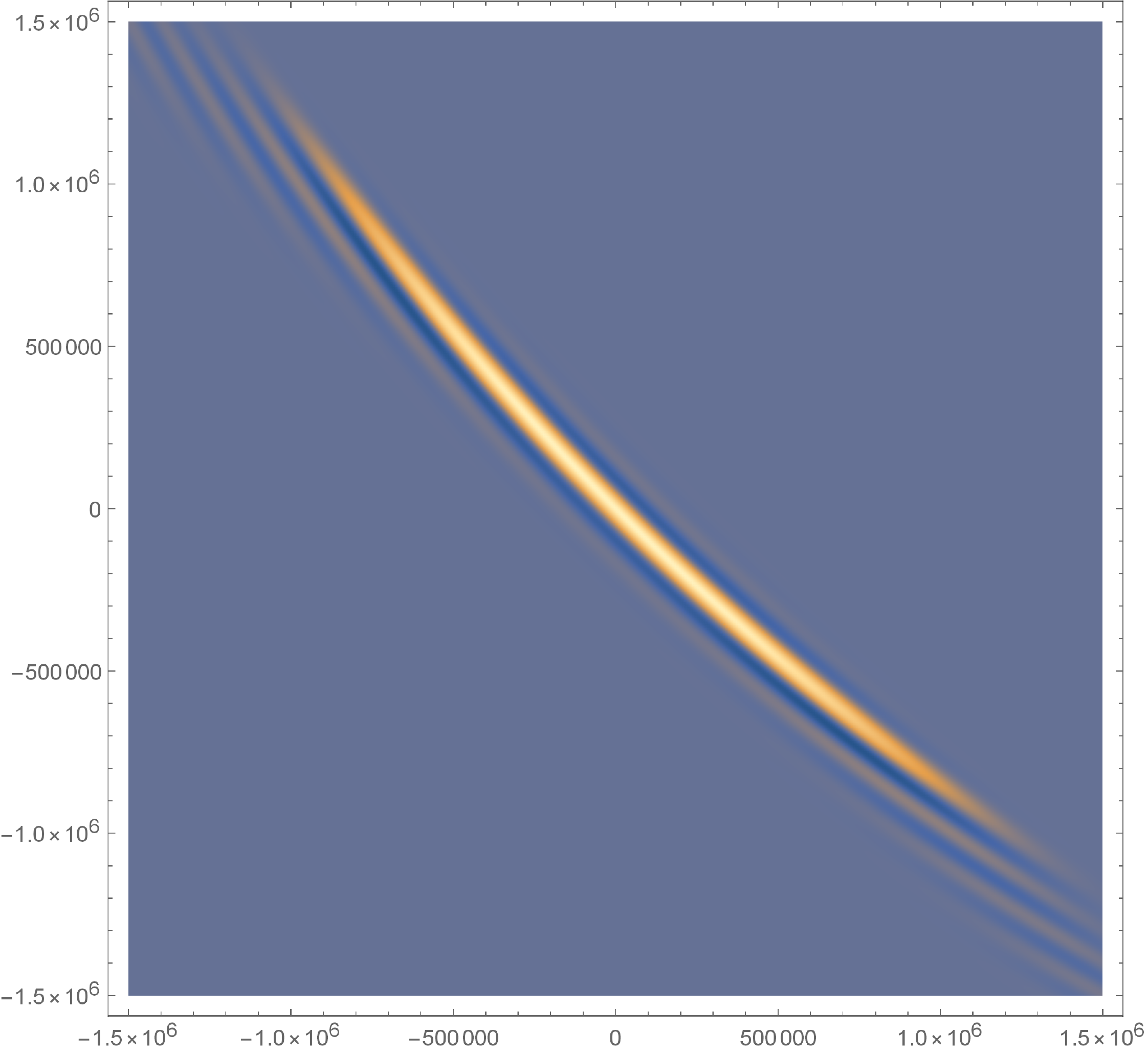}
\caption{A heat map visualizing the joint spectral distribution, $\mathcal{L}$. Each axis runs over the mode labels. More-strongly-coupled modes are lighter. Conservation of energy dictates that it is symmetric along the anti-diagonal axis and that the energy of down-converted modes are most likely to add up to the energy of the central frequency of the pump. The sinc-function-like behavior is due to the crystal having a finite physical extent (top-hat-like in the spatial domain therefore sinc-like in Fourier-transformed momentum domain). \label{Loz}}
\end{figure}

\noindent This represents a complicated network of two-mode squeezing interactions. This coupling matrix between the modes is visualized as a heat map in Figure \ref{Loz}.  

\begin{figure}[h!]
\centering
\includegraphics[scale=1.15]{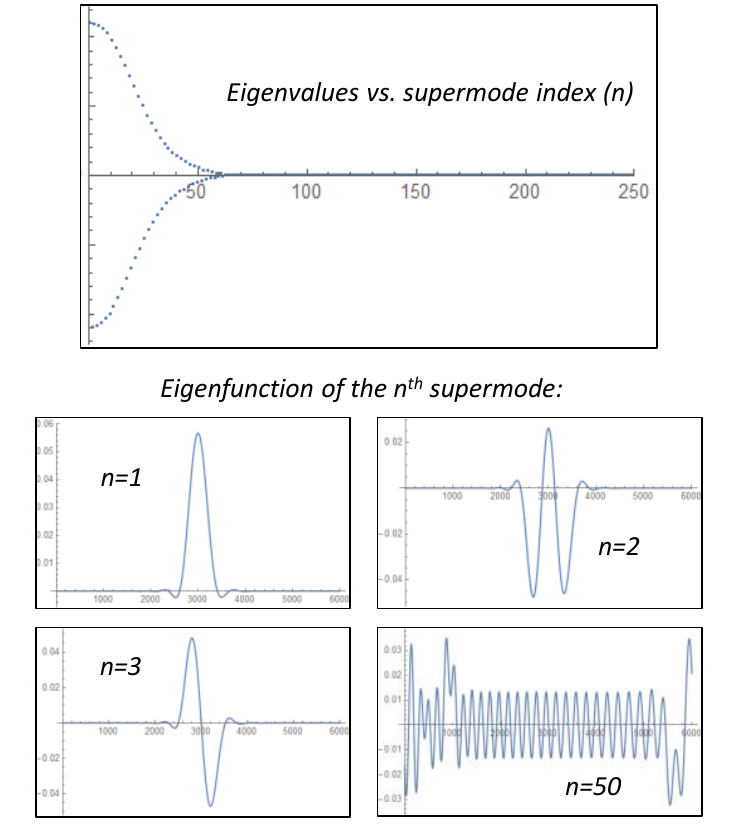}
\caption{A visual representation of a part of the supermode eigensystem we consider. Depicted are the first 250 eigenvalues, scaled to the largest value. Of special note is that after around mode 50 eigenvalues are neglibile, representing negligible squeezing so we therefore only consider the first 50 as contributing (all other assumed to be vacuum). Also shown are the first three eigenfunctions and the the last one we consider as contributing. \label{Eigen}}
\end{figure}

\noindent This matrix may be diagonalized yielding the supermode basis. In the supermode basis there is no coupling between modes and each squeezing transformation is single-mode. The eigenvalues of the eigensystem represent the squeezing parameter of the given supermode and the eigenvector gives the transformation between that supermode and the frequency basis. The supermodes are an entangled basis that mathematically simplify the description of the setup in several ways, but most importantly, the setup is sufficiently described by a much smaller number of mathematical modes. See Figure \ref{Eigen} for a graph of the eigenvalues and the first few eigenvectors of this system. For the parameters we consider, taking only the first 50 supermodes into account captures 98.75\% of all squeezing. Details are given in Ref.\cite{math}. Figure \ref{Eigen} also shows the last mode we consider.

We will consider the case where two photons are subtracted from the first supermode. The Wigner-function formalism will be advantageous for this calculation, so we will work in that picture. We start by writing the Wigner function of a high-dimensional (many-mode) Gaussian state. From Ref.\cite{raul} 

\begin{eqnarray}
W=\frac{1}{\pi^4\sqrt{\mathrm{Det}[\gamma]}}e^{-\textbf{r}^{T}\pmb{\gamma}^{-1}\textbf{r}},\label{wdef}
\end{eqnarray}    

\noindent where $\textbf{r}^{T}=[q_{1},p_{1},...,q_{f},p_{f}]$ is a vector containing all the quadrature variables ($p$ \& $q$) and $\pmb{\gamma}$ is a matrix defining their coupling. The coupling for a single-mode-squeezed state is 

\begin{eqnarray}
\pmb{\gamma}^{SMS}(s)=\left[
\begin{array}{cc}
e^{2s} & 0 \\
0 & e^{-2s}
\end{array}\right],
\end{eqnarray}  

\noindent where $s$ is the squeezing parameter. To represent the photon subtraction we will add two fictitious modes which are coupled to the first supermode via beam splitters (see inset in Fig.\ref{setup}). These modes start in the vacuum state and thus their $\pmb{\gamma}$ is simply the identity matrix of size four. The initial state of the source can thus be written as

\begin{eqnarray}
\pmb{\gamma}_{\mathrm{in}}=\textbf{I}_{4}\oplus\bigoplus_{j=1}^{50}\pmb{\gamma}^{SMS}_{j}(s_{j}),
\end{eqnarray}

\noindent where $s_{j}$ is the squeezing parameter of the $j^{\mathrm{th}}$ supermode. To represent the photon subtraction we will need the mixing and loss transformations are defined as in Ref.\cite{raul}

\begin{eqnarray}
\pmb{\gamma}&\rightarrow&\textbf{M}\pmb{\gamma}\textbf{M}^{T},\\
\pmb{\gamma}&\rightarrow&\textbf{L}\pmb{\gamma}\textbf{L}^{T}+\textbf{G}.
\end{eqnarray} 

\noindent These are a symplectic and completely-positive maps, respectively. For a beam splitter of transmissivity $T$ we have

\begin{eqnarray}
\textbf{M}(T)=\left[
\begin{array}{cccc}
\sqrt{T} & 0 & \sqrt{1-T} & 0 \\
0 & \sqrt{T} & 0 & \sqrt{1-T} \\
-\sqrt{1-T} & 0 & \sqrt{T} & 0 \\
0 & -\sqrt{1-T} & 0 & \sqrt{T}
\end{array}\right]
\end{eqnarray}

\noindent The first beam splitter couples the third mode of $\pmb{\gamma}_{\mathrm{in}}$ (i.e. the first supermode) with the first empty mode. The transmissivity is very high representing a weak beam splitter. The second beam splitter then couples the first and second modes and is 50-50. The matrices representing loss are given by Ref.\cite{raul}:

\begin{eqnarray}
\textbf{L}(\eta)&=&\eta\textbf{I},\label{loss1}\\
\textbf{G}(\eta)&=&(1-\eta)\textbf{I},\label{loss2}
\end{eqnarray}  

\noindent with $\eta$ the efficiency (which we will set to 1 for this initial analysis, representing lossless processes). The output coupling matrix before the heralded photon subtraction is then given by

\begin{eqnarray}
\pmb{\gamma}_{\mathrm{out}}(T,\eta)&=&\textbf{M}_{2}(0.5)\textbf{L}(\eta)\textbf{M}_{1}(T)\pmb{\gamma}_{\mathrm{in}}\nonumber \\
& &\times\textbf{M}_{1}^{T}(T)\textbf{L}^{T}(\eta)\textbf{M}_{2}^{T}(0.5)+\textbf{G}(\eta),\label{gout}
\end{eqnarray}

\noindent where the subscripts of the $\textbf{M}$ matrices indicate the first (weak) and second (50-50) beam-splitters. These matrices act as the identity on all other modes. The calculation of this matrix is given in Section II of Ref.\cite{math}.

The heralding event that signals that a two-photon subtraction has occurred is a detection in both of the two fictitious modes. The detectors are assumed to be of the threshold (on-off) type. The positive-operator-valued measure (POVM) for such a detection is 

\begin{eqnarray}
\hat{\Pi}=I-|0\rangle\langle 0|
\end{eqnarray}

\noindent To find the state after the heralded subtraction in the Wigner-function formalism we write

\begin{eqnarray}
W_{\mathrm{sub}}&=&(2\pi)^{2}\int_{-\infty}^{\infty}dq_{1}dp_{1}dq_{2}dp_{2}\nonumber \\ 
& &\times W_{\mathrm{out}}(q_{1},...,q_{n},p_{1},...,p_{n})\nonumber\\
& & \times W_{\hat{\Pi}_{A_{1}}}(q_{1},p_{1})W_{\hat{\Pi}_{A_{2}}}(q_{2},p_{2}),\label{wigsub}
\end{eqnarray}

\noindent where the labels $A_{1}$ and $A_{2}$ denote the first and second auxiliary modes (the first two modes). The Wigner function is given by Eq.(\ref{wdef}) in conjunction with Eq.(\ref{gout}). The Wigner function of the threshold (on-off) detector $\hat{\Pi}$ operator can be found with

\begin{eqnarray}
W_{\hat{\Pi}}&=&\frac{1}{2\pi}\int^{\infty}_{-\infty}dx\,e^{ixp}\left\langle q+\frac{x}{2}\right|\left(I-|0\rangle\langle 0|\right)\left|q-\frac{x}{2}\right\rangle,\nonumber \\
&=&\frac{1}{2\pi}\left(1-2\,e^{-p^{2}-q^{2}}\right).
\end{eqnarray}  

\noindent So now Eq.(\ref{wigsub}) can be written as

\begin{eqnarray}
W_{\mathrm{sub}}&=&\int_{-\infty}^{\infty}dq_{1}dp_{1}dq_{2}dp_{2} \,e^{-\textbf{r}^{T}\pmb{\gamma}_{\mathrm{out}}^{-1}\textbf{r}}\nonumber\\
& &\times\left(1-2\,e^{-p_{1}^{2}-q_{1}^{2}}-2\,e^{-p_{2}^{2}-q_{2}^{2}}\right.\nonumber \\
& &\quad +\left. 4\,e^{-p_{1}^{2}-q_{1}^{2}-p_{2}^{2}-q_{2}^{2}}\right)
\end{eqnarray}

\noindent where overall constants have been dropped (we'll re-normalize at the end). This can be more compactly written as 

\begin{eqnarray}
W_{\mathrm{sub}}=\int_{-\infty}^{\infty}dq_{1}dp_{1}dq_{2}dp_{2}\sum_{k=1}^{4}c_{k}e^{-\textbf{r}^{T}\pmb{\Gamma}_{k}\textbf{r}}\label{wsub}
\end{eqnarray}

\noindent Where $c_{1}=1$, $c_{2}=c_{3}=-2$, $c_{4}=4$, $\pmb{\Gamma}_{1}=\pmb{\gamma}_{\mathrm{out}}^{-1}$, $\pmb{\Gamma}_{2}=\pmb{\gamma}_{\mathrm{out}}^{-1}+\textbf{I}_{A_{1}}\oplus\pmb{\emptyset}_{102}$, $\pmb{\Gamma}_{3}=\pmb{\gamma}_{\mathrm{out}}^{-1}+\textbf{I}_{A_{2}}\oplus\pmb{\emptyset}_{102}$, $\pmb{\Gamma}_{3}=\pmb{\gamma}_{\mathrm{out}}^{-1}+\textbf{I}_{A_{1}}\oplus\textbf{I}_{A_{2}}\oplus\pmb{\emptyset}_{100}$, and $\pmb{\emptyset}_{n}$ represents a matrix of all zeros of size $n$.\\

\noindent To perform this integration we need to subdivide the symmetric matrices $\pmb{\Gamma}_{k}$

\begin{eqnarray}
\pmb{\Gamma}_{k}=\left[
\begin{array}{c|c}
\pmb{\Gamma}^{A}_{k} & \begin{matrix} & \pmb{\sigma}_{k} & \end{matrix} \\
\hline		
\begin{matrix}  \\ \pmb{\sigma}^{T}_{k} \\ \, \end{matrix} & \pmb{\Gamma}^{S}_{k}
\end{array}
\right].\label{cov1}
\end{eqnarray}

\noindent The matrices $\pmb{\Gamma}^{A}_{k}$ will contain the auxiliary modes, which are the ones that will need to be integrated out; they are $4\times 4$. The matrices $\pmb{\Gamma}^{S}_{k}$ contain the supermodes that will remain; they are $100 \times 100$. The matrices $\pmb{\sigma}_{k}$ are $4\times 100$. We also subdivide the vector of quadrature variables as $\textbf{r}^{T}=\left[\textbf{U}^{T}, \textbf{V}^{T}\right]$, where $\textbf{U}$ contains the four variables to be integrated over and $\textbf{V}$ contains the remainder. So,

\begin{eqnarray}
\textbf{r}^{T}\pmb{\Gamma}_{k}\textbf{r}=\textbf{U}^{T}\pmb{\Gamma}^{A}_{k}\textbf{U}+\textbf{V}^{T}\pmb{\Gamma}^{S}_{k}\textbf{V}+2\textbf{V}^{T}\pmb{\sigma}^{T}_{k}\textbf{U}.\label{cov2}
\end{eqnarray}

\noindent Now, we remark that

\begin{eqnarray}
& &\textbf{U}^{T}\pmb{\Gamma}^{A}_{k}\textbf{U}+\textbf{V}^{T}\pmb{\Gamma}^{S}_{k}\textbf{V}+2\textbf{V}^{T}\pmb{\sigma}^{T}_{k}\textbf{U}\label{cov3}\\
& &\quad=\left(\textbf{U}+\pmb{\Gamma}^{A^{-1}}_{k}\pmb{\sigma}_{k}\textbf{V}\right)^{T}\pmb{\Gamma}^{A}_{k}\left(\textbf{U}+\pmb{\Gamma}^{A^{-1}}_{k}\pmb{\sigma}_{k}\textbf{V}\right)\nonumber \\
& & \quad \quad + \textbf{V}^{T}\left(\pmb{\Gamma}^{S}_{k}-\pmb{\sigma}_{k}^{T}\pmb{\Gamma}^{A^{-1}}_{k}\pmb{\sigma}_{k}\right)\textbf{V}.\nonumber
\end{eqnarray}

\noindent We then perform a change of variables as $\textbf{U}'=\left(\textbf{U}+\pmb{\Gamma}^{A^{-1}}_{k}\pmb{\sigma}_{k}\textbf{V}\right)$, the determinant of this Jacobian is one and the integral over the $\textbf{U}'$ can proceed as normal. So Eq.(\ref{wsub}) becomes

\begin{eqnarray}
W_{\mathrm{sub}}=\sum_{k=1}^{4}\frac{c_{k}}{\sqrt{\mathrm{Det}\left[\pmb{\Gamma}^{A}_{k}\right]}}e^{-\textbf{V}^{T}\left(\pmb{\Gamma}^{S}_{k}-\pmb{\sigma}_{k}^{T}\pmb{\Gamma}^{A^{-1}}_{k}\pmb{\sigma}_{k}\right)\textbf{V}}.\label{cov4}
\end{eqnarray}

\begin{figure}[h!]
 \centering
  \includegraphics[scale=0.73]{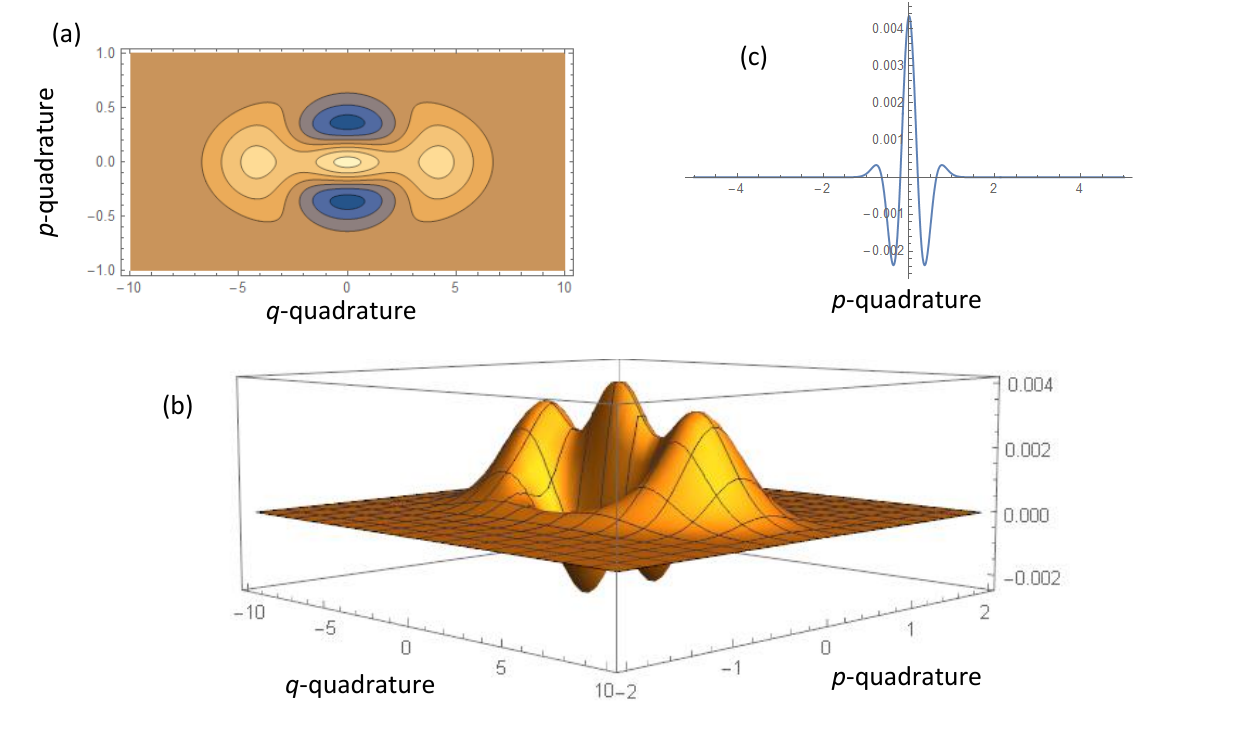}
 \caption{The Wigner function after photon subtraction. (a) is a contour plot, (b) is a 3-D plot, and (c) is a slice though the $p$ quadrature. \label{wig1}}
 \end{figure}

\noindent This calculation is in Section III of Ref.\cite{math}. Figure \ref{wig1} contains plots of the first supermode after photon-subtraction.

Returning to Fig.\ref{setup} we see that the next mathematical transformation we need to make is from the supermode basis to the pixel basis (via the frequency basis). The weight of the conversion between an individual supermode and a pixel mode is given by the normalized integral of the eigenfunction of that supermode over the frequency domain corresponding to the given pixel. These eigenfunctions are the ones we show samples of in Fig.\ref{Eigen}. Each mode is modulated with a Gaussian to represent being mixed with the homodyne reference field. The reference field may be shaped in many other ways, but we take this simple case. The coarse-graining of the frequency basis into the pixel basis is done evenly (i.e. the integrations are taken over domains of the same size.) Section IV of the Mathematica$^{\mathrm{TM}}$ file in Ref.\cite{math} has the details of that calculation, and Section V concerns the application of detection loss to the pixel modes \--- in the same manner as in Eqs.(\ref{loss1},\ref{loss2}).

There is also a ``switching matrix'' that puts the $p$'s and $q$'s next to each other so that the integration can be written more easily. We include this matrix here as we use a different one than in Ref.\cite{raul}. That is

\begin{eqnarray}
\left[\begin{array}{c}
p_{1} \\ p_{2} \\ q_{1} \\ q_{2} 
\end{array}\right] =
\left[\begin{array}{cccc}
0 & 1 & 0 & 0 \\
0 & 0 & 0 & 1 \\
1 & 0 & 0 & 0 \\
0 & 0 & 1 & 0
\end{array}\right]
\left[\begin{array}{c}
q_{1} \\ p_{1} \\ q_{2} \\ p_{2} 
\end{array}\right]. 
\end{eqnarray}  

Then, to represent the fact that we are dividing the detection outcomes into two half-planes whose line of division sits at angle (with respect to the horizontal in quadrature space) equal to the measurement setting, we perform a final transformation by rotating the function using the standard rotation matrices. This new Wigner function is $W_{BD}(q_{1},q_{2},p_{1},p_{2})$. Then we integrate the $p$ modes over \emph{all} space (as it will be the $q$ quadratures that are split up to form the dichotomic observables). So we have

\begin{flalign}
& \quad \quad W_{\mathrm{f}}\left(\phi_{a_{j}},\phi_{b_{k}},q_{1},q_{2}\right)\label{wf} && \\
& \quad \quad \quad \quad =\int_{-\infty}^{\infty}\int_{-\infty}^{\infty}dq_{1}dq_{2}W_{\mathrm{BD}}(q_{1},q_{2},p_{1},p_{2}).\nonumber &&
\end{flalign}

\noindent The same change of variables as in Eqs.(\ref{cov1},\ref{cov2},\ref{cov3},\ref{cov4}) is utilized. This leaves only a distribution over the $q$ modes.

The dichotomic observable of the homodyne Bell test is whether the field is measured to be in one or the other sides of the half-plane of quadrature space (as defined by a choice of bias phase during homodyne) for either party. One half (the positive) of the plane is assigned the value $+1$ and the other is assigned value $-1$ for each measurement event (heralded by the photon subtraction event) and party. The $(q_{1},q_{2})$ plane is divided up into four quadratures by whether $q_{1}$ and $q_{2}$ are less than or more than zero. That is, possible results are $+1=(+1)_{a}(+1)_{b}$, $+1=(-1)_{a}(-1)_{b}$, $-1=(-1)_{a}(+1)_{b}$, and $-1=(+1)_{a}(-1)_{b}$; where the subscripts indicate party. The correlator is then the probability of these various outcomes times the total outcome. For the case of a positive outcome we have  

\begin{flalign}
& \quad  P_{+1}\left(\phi_{a_{j}},\phi_{b_{k}}\right)&& \\
& \quad \quad \quad \quad =\int_{0}^{\infty}\int_{0}^{\infty}dq_{1}dq_{2}\,W_{\mathrm{f}}\left(\phi_{a_{j}},\phi_{b_{k}},q_{1},q_{2}\right)\nonumber &&\\
& \quad \quad\quad \quad \quad \quad +\int^{0}_{-\infty}\int^{0}_{-\infty}dq_{1}dq_{2}\,W_{\mathrm{f}}\left(\phi_{a_{j}},\phi_{b_{k}},q_{1},q_{2}\right)\nonumber &&\\
& \quad \quad \quad \quad =\int_{0}^{\infty}\int_{0}^{\infty}dq_{1}dq_{2}\,W_{\mathrm{f}}\left(\phi_{a_{j}},\phi_{b_{k}},q_{1},q_{2}\right)\nonumber &&\\
& \quad \quad\quad \quad \quad \quad +\int_{0}^{-\infty}\int_{0}^{-\infty}dq_{1}dq_{2}\,W_{\mathrm{f}}\left(\phi_{a_{j}},\phi_{b_{k}},q_{1},q_{2}\right)\nonumber &&\\
&\quad \quad \quad \quad  =\int_{0}^{\infty}\int_{0}^{\infty}dq_{1}dq_{2}\,W_{\mathrm{f}}\left(\phi_{a_{j}},\phi_{b_{k}},q_{1},q_{2}\right)\nonumber &&\\
& \quad \quad\quad \quad \quad \quad  +\int_{0}^{\infty}\int_{0}^{\infty}dq_{1}dq_{2}\,W_{\mathrm{f}}\left(\phi_{a_{j}},\phi_{b_{k}},-q_{1},-q_{2}\right)\nonumber &&\\
& \quad \quad \quad \quad  =2\int_{0}^{\infty}\int_{0}^{\infty}dq_{1}dq_{2}\,W_{\mathrm{f}}\left(\phi_{a_{j}},\phi_{b_{k}},q_{1},q_{2}\right),\nonumber &&
\end{flalign} 

\noindent where $W_{\mathrm{f}}$ is the final Wigner function given by Eq.(\ref{wf}). In the second equality the limits of two integrals are flipped so the negative signs cancel, and the last equality is due to the fact that the Wigner function is symmetric: $W_{\mathrm{f}}\left(q_{1},q_{2}\right)=W_{\mathrm{f}}\left(-q_{1},-q_{2}\right)$. The probability for a negative outcome is simply

\begin{flalign}
& \quad P_{-1}\left(\phi_{a_{j}},\phi_{b_{k}}\right)\nonumber && \\
&\quad\quad =1-P_{+1}\left(\phi_{a_{j}},\phi_{b_{k}}\right)&&\nonumber\\
&\quad\quad =1-2\int_{0}^{\infty}\int_{0}^{\infty}dq_{1}dq_{2}\,W_{\mathrm{f}}\left(\phi_{a_{j}},\phi_{b_{k}},q_{1},q_{2}\right).&&
\end{flalign}

\noindent And the correlator is given as

\begin{eqnarray}
E\left(\phi_{a_{j}},\phi_{b_{k}}\right)&=&+1\,P_{+1}\left(\phi_{a_{j}},\phi_{b_{k}}\right)-1\,P_{-1}\left(\phi_{a_{j}},\phi_{b_{k}}\right)\nonumber\\
&=&4\int_{0}^{\infty}\int_{0}^{\infty}dq_{1}dq_{2}\,W_{\mathrm{f}}\left(\phi_{a_{j}},\phi_{b_{k}},q_{1},q_{2}\right)-1,\nonumber
\end{eqnarray}

\noindent which can be used the calculate the two-party Bell parameter in the CHSH inequality (Eq.\ref{chsh}), $\mathcal{S}_{2}$:

\begin{eqnarray}
\mathcal{S}_{2}=\frac{1}{2}\left|E \left(\phi_{a_{1}},\phi_{b_{1}}\right)\right.&+& E\left(\phi_{a_{1}},\phi_{b_{2}}\right)\\
 &+& \left. E\left(\phi_{a_{2}},\phi_{b_{1}}\right)-E\left(\phi_{a_{2}},\phi_{b_{2}}\right)\right|,\nonumber
\end{eqnarray}

\noindent where $\mathcal{S}_{2}\ > 1$ indicates violation of local-realistic theories under the assumptions implicit in CHSH.

These calculations are carried out by the symbolic program Mathematica$^{\mathrm{TM}}$. The integrations are done both analytically and numerically as a check on correctness. Carrying out multidimensional integrals over the full plane is relatively easy (via the method described in earlier in this Section), but such integrals over the half-plane are much more challenging. Therefore these final integrations are very time-consuming to solve and this creates a computational bottleneck that is very difficult to work around.  This will be discussed further in the next couple Sections. 

The above calculation must be performed numerically for every possible combination of measurement choices (the $\phi$'s) in order to find which choices of measurement angles yield the highest violation. Since continuous results are not available (the time-intensive integration must be carried out for each combination) enough points must be chosen so that we can be reasonably sure of local extrema. Since the bell value is not necessarily symmetric we also allow one party a total global rotation on all measurement directions. 

In Section VI of \cite{math} we find the maximum violation of this inequality to be 1.032. This is a small improvement over the result of the setup used in Ref.\cite{raul} which was 1.024. This is surprising as it is likely that much of the correlation is lost in the conversion to pixel basis. 

The central premise of this paper has been proven \--- that the setup as described, in principle, has the power to violate a local-realistic inequality. However it is likely still too small to be technologically useful with the setup in the configuration as modeled. Possible modifications and the challenges inherent in extending the calculation are further discussed in the following Section.     

\section{Four-Party Case}

The power of the setup described here is its ability to spread stronger-than-classical correlations over very many physically distinct modes. We can model this by simply decreasing the granularity of the pixel-binning. In this section we consider four pixel modes which would represent four parties attempting to violate a Bell-type inequality. The reason we go from two to four is that we know from the analysis in Ref.\cite{Acin} that the parity of the number of parties/modes must match the parity of the photon-number subtraction, therefore since we consider a two-photon subtraction using four parties is the next logical choice.

The calculation in Section V must be re-done to account for the larger number of modes, however most of these modifications are rather direct, and result in very-large equations, so we do not report that calculation here. The results are detailed in the Mathematica$^{\mathrm{TM}}$ file \--- Section VII of \cite{math}.  

However, a significant difference between the two-party and four-party case is the inequalities that must be used. We require an inequality that accommodates larger numbers of parties. There are several classes of such inequalities and, unlike in the two-party case, in the four-party case it is not clear which are optimal, so we try several. 

We start by utilizing the recursion method detailed in section 5.2 of ``Interference of Light and Bell's Theorem'' by Belinski\u{\i} and Klyschko, Ref.\cite{BK}. This method generates inequalities for any party number. For the four-party case the Bell value for the Belinski\u{\i} and Klyschko (BK) Inequality (called the Mermin-Klyschko Inequality \cite{M}) is given by

\begin{eqnarray}
\mathcal{BK}_{4}&=&\frac{1}{4}\left[E_{4}(\phi_{a1}, \phi_{b1}, \phi_{c1}, \phi_{d1})+ E_{4}(\phi_{a2},\phi_{b1}, \phi_{c1}, \phi_{d1})\right. \nonumber \\
 & &+E_{4}(\phi_{a1}, \phi_{b2}, \phi_{c1}, \phi_{d1})-E_{4}(\phi_{a2},\phi_{b2}, \phi_{c1}, \phi_{d1}) \nonumber \\
 & &+E_{4}(\phi_{a1}, \phi_{b1},\phi_{c2}, \phi_{d1})-E_{4}(\phi_{a2}, \phi_{b1}, \phi_{c2},\phi_{d1}) \nonumber \\
 & &-E_{4}(\phi_{a1}, \phi_{b2}, \phi_{c2}, \phi_{d1}) -E_{4}(\phi_{a2}, \phi_{b2}, \phi_{c2}, \phi_{d1})\nonumber \\
 & &+E_{4}(\phi_{a1}, \phi_{b1}, \phi_{c1}, \phi_{d2})-E_{4}(\phi_{a2}, \phi_{b1}, \phi_{c1}, \phi_{d2})\nonumber \\
 & &-E_{4}(\phi_{a1}, \phi_{b2}, \phi_{c1}, \phi_{d2})-E_{4}(\phi_{a2}, \phi_{b2}, \phi_{c1}, \phi_{d2})\nonumber \\
 & &-E_{4}(\phi_{a1}, \phi_{b1},\phi_{c2},\phi_{d2})-E_{4}(\phi_{a2}, \phi_{b1}, \phi_{c2}, \phi_{d2})\nonumber \\
 & &\left. -E_{4}(\phi_{a1}, \phi_{b2}, \phi_{c2},\phi_{d2})+E_{4}(\phi_{a2}, \phi_{b2}, \phi_{c2},\phi_{d2})\right].\nonumber
\end{eqnarray}

Each party has a choice of two measurement angles, as in the two-party case.

For even party number the well-known Svetlichny Inequality \cite{S} is also equivalent to the above inequality. However the Mermin Inequality differs. For the four-party case it is given as

\begin{eqnarray}
\mathcal{M}_{4}&=&\frac{1}{4}\left[E_4(\phi_{a2}, \phi_{b1}, \phi_{c1}, \phi_{d1})+ E_4(\phi_{a1}, \phi_{b2}, \phi_{c1}, \phi_{d1})\right. \nonumber \\ 
& &+E_4(\phi_{a1}, \phi_{b1}, \phi_{c2}, \phi_{d1})- E_4(\phi_{a2}, \phi_{b2}, \phi_{c2}, \phi_{d1})\nonumber \\
& &+ E_4(\phi_{a1}, \phi_{b1}, \phi_{c1}, \phi_{d2})- E_4(\phi_{a2}, \phi_{b2}, \phi_{c1}, \phi_{d2})\nonumber \\
& &- \left. E_4(\phi_{a2}, \phi_{b1}, \phi_{c2}, \phi_{d2}) - E_4(\phi_{a1}, \phi_{b2}, \phi_{c2}, \phi_{d2})\right].\nonumber
\end{eqnarray}

For both inequalities violation occurs if the Bell value is less than negative one or greater than one. 

As before, all angles for a given party are given the freedom to be rotated locally \--- apart from the first party. For the sake of neatness and clarity this is not included in the above equations.

Another notable difference is that the integrations over the half-plane Wigner functions are far more computationally intensive to perform. This exasperates a problem that was also present in the two-party case: that if we wish to modify the setup (in what ways this could be done is discussed later in this section) in an attempt to optimize the Bell violation we must re-do the calculation for every possible combination of parameters. 

We find a maximum value of 0.757, which of-course is not even near the edge of the quantum-classical divide. The parameters yielding maximum violation are given in Ref.\cite{math}.

So far we have been making the simplest and most sensible-seeming choices for the details of the setup based on a specific experimental implementation \cite{Fabre2} (all numerical values are taken directly from that experiment and we make choices for things like homodyne shape, mirror transmissivity, loss, etc. that are experimentally reasonable, lead to simplifications, or would seem, intuitively, to make violation more likely). However, the experimental setup itself is highly adjustable and many variations might be employed, and it is not necessarily the case that the choices we have made are even close to being the best.   

Some examples of how the setup could be adjusted include: taking fuller advantage of the photon-subtraction procedure's ability to fully shape the spectral mode of the subtraction, employing photon addition, changing the shape of the homodyne envelope, changing the domains of the pixel modes, trying 4-photon subtraction, and utilizing more specifically-tailored inequalities \--- including inequalities specifically for continuous-variable situations. In fact the pump spectral profile itself may also be shaped and this has already been studied as an avenue for optimizing other applications \cite{AT}. It may also be the case that similar setups utilizing different geometries and different degrees of freedom might be more efficient at violating inequalities \cite{furusawa,pfister2,huntington}.   

The issue again is that there is a very difficult computational bottleneck at the last stage of the calculation. Meaning that, as the calculation procedure is now, any adjustment could result in days of extra computation on a standard modern desktop. Searching a parameter space as large as is available would be near-impossible as-is. It is also important to point out that similar optimization may be attempted for the two-party case, the computational bottle neck exists there as well, but is significantly less severe. 

However, it is very likely the calculation could be modified to run more quickly and employ larger computational resources, which is why we make it publicly available at \cite{math}. 

\section{Summary and Conclusions}

We have built a model of a synchronously-pumped optical parametric oscillator which has undergone photon subtraction, and investigated its ability to violate a Bell-type inequality using highly-efficient homodyne detection. The inequality is constructed by discretizing quadrature space to yield the dichotomic observables most such inequalities require. The number of output modes \--- or parties \--- can be controlled by choosing how coarse-grained to make the division of the frequency comb into frequency-based pixel modes, which \--- in principle \--- could be in the millions. 

The advantage of this type of setup is twofold: highly-efficient homodyne detection may be employed, and very many physical modes are available for quantum-technological tasks. 

We studied two different possibilities, a two-party case and a four-party case. Quite surprisingly we found an increase in violation for the two-party case over simpler setups \--- though the amount of violation is still likely too small to be easily experimentally observable. This constitutes a proof-of-principle of the non-classically of the system in question. 

We had expected that the degree of violation would grow with the number of parties, but when the four-party case is examined we find a reduction of the Bell parameter to below the quantum limit. Our conjecture is that this lack of violation in the four-party case is due to the fact that the parameter space of possible experimental configurations is both massive \--- and computationally difficult to search \--- and that there are likely parameter combinations which would lead to violation. In the hopes that others will try to simplify this calculation, or employ more robust computational resources, and search this parameter space for violations, we make all code and calculations publicly available at \cite{math}.     

\section*{Acknowledgments}

\noindent This work is supported by the European Union Grant QCUMbER (No. 665148), the ERC Advanced Grant QIT4QAD, and the French National Research Agency through the project COMB (grant number ANR-13-BS04-0014). Discussions with Elham Kashefi and Claude Fabre were invaluable.


\begin{thebibliography}{10}

\bibitem{BellNL}``Bell nonlocality'', N. Brunner, D. Cavalcanti, S. Pironio, V. Scarani, and S. Wehner, Rev. Mod. Phys. \textbf{86}, 419 (2014).

\bibitem{met}``Quantum Metrology'', V. Giovannetti, S. Lloyd, and L. Maccone, Phys. Rev. Lett. \textbf{96}, 010401 (2006).

\bibitem{met2}``Quantum Sensing'', C. L. Degen, F. Reinhard, and P. Cappellaro, Rev. Mod. Phys. \textbf{89}, 035002 (2017).

\bibitem{crypt}``Quantum cryptography'', N. Gisin, G. Ribordy, W. Tittel, and H. Zbinden, Rev. Mod. Phys. \textbf{74}, 145 (2002).

\bibitem{qc}``Quantum Computation and Quantum Information'', Cambridge University Press (2010).

\bibitem{MBQC}``One-way Quantum Computation \--- a tutorial introduction'', D. Browne, and H. Briegel, arXiv:quant-ph/0603226v2 (2006).

\bibitem{CVMBQC}``Quantum Computing with Continuous-Variable Clusters'', M. Gu, C. Weedbrook, N.C. Menicucci, T.C. Ralph, and P. van Loock, Phys. Rev. A \textbf{79}, 062318 (2009).

\bibitem{Fabre1}``Multimode squeezing of frequency combs'', G.J. de Valc\'{a}rcel, G. Patera, N. Treps, and C. Fabre, Phys. Rev. A Rap. Comm. \textbf{74}, 061801(R) (2006).

\bibitem{pfister1}``One-Way Quantum Computing in the Optical Frequency Comb'', N.C. Menicucci, S.T. Flammia, and O. Pfister, Phys. Rev. Lett. \textbf{101}, 130501 (2008).

\bibitem{Fabre2}``Generation and Characterization of Multimode Quantum Frequency Combs'', O. Pinel, P. Jian, R.M. de Ara\'{u}jo, J. Feng, B. Chalopin, C. Fabre, and N. Treps, Phys. Rev. Lett. \textbf{108}, 083601 (2012).

\bibitem{furusawa}``Ultra-large-scale continuous-variable cluster states multiplexed in the time domain'', S. Yokoyama, R. Ukai, S.C. Armstrong, C. Sornphiphatphong, T. Kaji, S. Suzuki, J. Yoshikawa, H. Yonezawa, N.C. Menicucci, and A. Furusawa, Nat. Phot. \textbf{7}, 982 (2013).

\bibitem{pfister2}``Experimental realization of multipartite entanglement of 60 modes of a quantum optical frequency comb'', M. Chen, N.C. Menicucci, and O. Pfister, Phys. Rev. Lett. \textbf{112}, 120505 (2014).

\bibitem{huntington}``Quantum teleportation in space and frequency using entangled pairs of photons from a frequency comb'', H. Song, H. Yonezawa, K.B. Kuntz, M. Heurs, E.H. Huntington, Phys. Rev. A \textbf{90}, 042337 (2014).

\bibitem{qpg1}``A quantum pulse gate based on spectrally engineered sum frequency generation'', A. Eckstein, B. Brecht and C. Silberhorn, Opt. Express \textbf{15}, 13770 (2011).

\bibitem{qpg2}``Multimode theory of single-photon subtraction'', V.A. Averchenko, C. Jacquard, V. Thiel, C. Fabre and N. Treps, New J. Phys. \textbf{18}, 083042 (2016).

\bibitem{qpg3}``Tomography of a Mode-Tunable Coherent Single-Photon Subtractor'', Y.S. Ra, C. Jacquard, A. Dufour, C. Fabre, and N. Treps, Phys. Rev. X \textbf{7}, 031012 (2017).

\bibitem{patera}``Quantum theory of synchronously pumped type I optical parametric oscillators: characterization of the squeezed supermodes'', G. Patera, N. Treps, C. Fabre, and G.J. de Valc\'{a}rcel, Eur. Phys. J. D \textbf{56}, 123 (2010).

\bibitem{ferrini}``Compact Gaussian quantum computation by multi-pixel homodyne detection'',  G. Ferrini, J.P. Gazeau, T. Coudreau, C. Fabre, and N. Treps, New J. Phys. \textbf{15}, 093015 (2013).

\bibitem{trepsnat} ``Wavelength-multiplexed quantum networks with ultrafast frequency combs'' J. Roslund, R.M. de Ara\'{u}jo, S. Jiang, C. Fabre, and N. Treps, Nat. Phot. \textbf{8}, 109 (2014).

\bibitem{witness}``Full multipartite entanglement of frequency comb Gaussian states'' S. Gerke, J. Sperling, W. Vogel, Y. Cai, J. Roslund, N. Treps, and C. Fabre, Phys. Rev. Lett. \textbf{114}, 050501 (2015).

\bibitem{opt}``Optimization strategies in measurement based quantum computation'', G. Ferrini, J. Roslund, F. Arzani, Y. Cai, C. Fabre, and N. Treps, $<$hal-01026145v1$>$ (2014).

\bibitem{AT}``Versatile engineering of multimode squeezed states by optimizing the pump spectral profile in spontaneous parametric down-conversion'', F. Arzani, C. Fabre, and N. Treps, arXiv:1709.10055 (2018).

\bibitem{Acin}``Tests of multimode quantum nonlocality with homodyne measurements'', A. Ac\'{i}n, N.J. Cerf, A. Ferraro, and J. Niset, Phys. Rev. A \textbf{79}, 012112 (2009).

\bibitem{raul}``Loophole-free test of quantum nonlocality using high-efficiency homodyne detectors'', R. Garc\'{i}a-Patr\'{o}n, J. Fiur\'{a}\v{s}ek, and N.J. Cerf, Phys. Rev. A \textbf{71}, 022105 (2005).

\bibitem{raulL}``Proposal for a loophole-free Bell test using homodyne detection'', R. Garc\'{i}a-Patr\'{o}n, J. Fiur\'{a}\v{s}ek, N. J. Cerf, J. Wenger, R. Tualle-Brouri, and Ph. Grangier, Phys. Rev. Lett. \textbf{93}, 130409 (2004).

\bibitem{chsh}``Proposed experiment to test local hidden-variable theories'', J.F. Clauser, M.A. Horne, A. Shimony, R.A. Holt, Phys. Rev. Lett., \textbf{23}, 15 (1969).

\bibitem{math}``Online Mathematica file'' to be added later.

\bibitem{BK}``Interference of Light and Bell's Theorem'', A.V. Belinski\u{\i} and D.N. Klyschko, Phys. Usp. \textbf{36}, 653 (1993).

\bibitem{M}``Simple unified form for the major no-hidden-variables theorems'', N. D. Mermin, Phys. Rev. Lett. \textbf{65}, 1838 (1990). 

\bibitem{S}``Distinguishing three-body from two-body nonseparability by a Bell-type inequality'', G. Svetlichny, Phys. Rev. D \textbf{35}, 3066 (1987).

\end{thebibliography}
\end{document}